\documentclass{jfm}
\usepackage{graphicx}
\usepackage{epstopdf, epsfig}

\usepackage[colorlinks, allcolors=blue]{hyperref} 

\shorttitle{Two-layer Thermally Driven Turbulence: Mechanisms for Interface Breakup}
\shortauthor{H.-R. Liu, K. L. Chong, Q. Wang, C. S. Chong, R. Verzicco and D. Lohse}

\title{Two-layer Thermally Driven Turbulence: \\Mechanisms for Interface Breakup}

\author{Hao-Ran Liu\aff{1},
  Kai Leong Chong\aff{1},
  Qi Wang\aff{1,2},
  Chong Shen Ng\aff{1},
  Roberto Verzicco\aff{3,4,1}
 \and Detlef Lohse\aff{1,5,}\corresp{\email{d.lohse@utwente.nl}}}

\affiliation{\aff{1}Physics of Fluids Group and Max Planck Center Twente for Complex Fluid Dynamics,\\
MESA+Institute and J. M. Burgers Centre for Fluid Dynamics, University of Twente,\\
P.O. Box 217, 7500AE Enschede, The Netherlands
\aff{2}Department of Modern Mechanics, University of Science and Technology of China, Hefei 230027, China
\aff{3}Dipartimento di Ingegneria Industriale, University of Rome ``Tor Vergata", Via del Politecnico 1, Roma 00133, Italy
\aff{4}Gran Sasso Science Institute - Viale F. Crispi, 7 67100 L’Aquila, Italy
\aff{5}Max Planck Institute for Dynamics and Self-Organization, Am Fassberg 17, 37077 Göttingen, Germany}

\begin{document}

\maketitle

\begin{abstract}
It is commonly accepted that the breakup criteria of drops or bubbles in turbulence is governed by surface tension and inertia. However, also {\it{buoyancy}} can play an important role at breakup. In order to better understand this role, here we numerically study two-dimensional Rayleigh-B\'enard convection for two immiscible fluid layers, in order to identify the effects of buoyancy on interface breakup. We explore the parameter space spanned by the Weber number $5\leq We \leq 5000$ (the ratio of inertia to surface tension) and the density ratio between the two fluids $0.001 \leq \Lambda \leq 1$, at fixed Rayleigh number $Ra=10^8$ and Prandtl number $Pr=1$. At low $We$, the interface undulates due to plumes. When $We$ is larger than a critical value, the interface eventually breaks up. Depending on $\Lambda$, two breakup types are observed: The first type occurs at small $\Lambda \ll 1$ (e.g. air-water systems) when local filament thicknesses exceed the Hinze length scale. The second, strikingly different, type occurs at large $\Lambda$ with roughly $0.5 < \Lambda \le 1$ (e.g. oil-water systems): The layers undergo a periodic overturning caused by buoyancy overwhelming surface tension. For both types the breakup, criteria can be derived from force balance arguments and show good agreement with the numerical results.
\end{abstract}


\section{Introduction}
Liquid can break up or fragment in multiphase turbulence \citep{nature02, frag, chao, frag2}. This physical phenomenon is very important for raindrops \citep{rain, zaleski}, for ocean waves and the resulting spray \citep{spray}, and even for the transmission of virus-laden droplets during coughing or sneezing \citep{covid,covid-steven}. Once the physical mechanisms governing this important phenomenon are understood, one can deduce quantitative criteria for the breakup to occur. In turbulence, for drops or bubbles the breakup criteria can be deduced from the balance of inertial and surface tension forces, as developed in the Kolmogorov-Hinze theory \citep{kol,hinze}. This theory well predicts the breakup criteria in experimental and numerical results in various flow systems, e.g., homogeneous isotropic turbulence \citep{hit99exp,bt,LB19JFM}, shear flows \citep{luka19jfm}, pipe flows \citep{pipe} and ocean waves \citep{nature02, pop16}.

Whilst the classical Kolmogorov-Hinze theory considers only surface tension and inertial forces, in many multiphase turbulent flows also {\it buoyancy} can play an important role. Examples of multiphase buoyant turbulence include the hotspots and superswells in Earth's mantle \citep{nature99, science} and even flows during sneezing and exhalation \citep{sneezing}. In such flows, the breakup of the interface between the fluids is the key phenomenon. Yet, the exact mechanisms that drives interface breakup when buoyancy is crucial is unknown.

The objective of the present work is to shed light on this mechanism. As examples for turbulent flow where buoyancy is important and at the same time can easily be tuned, we take thermal convection, namely Rayleigh-B\'enard (RB) convection \citep{rev1,rev2,chilla} of two immiscible fluids. We numerically investigate the breakup mechanisms of the interface between the two immiscible fluids. The immiscible fluids are first arranged in two layers according to their densities and then heated from below and cooled from above. Most previous studies of such two-layer RB convection were conducted in the non-turbulent regime \citep{low1,low2,low3,low-ra}. Experimental studies in the turbulent regime 
were reported by \cite{xie}, who focused on the flow structures in each layer and the coupling modes between the flows in the two layers, including viscous coupling and thermal coupling. Besides the classical rectangular/cylindrical configuration, the two-layer RB convection in spherical-shell geometry was also numerically studied by \cite{earth}. However, these previous studies only considered the case for strong surface tension, where the interface between the layers does not break up. In this study we will for the first time explore the case with interface breakup, which happens when surface tension is sufficiently small.

The control parameters of two-layer RB convection are the density ratio $\Lambda$ between two fluids and the Weber number $We$, which is the ratio of inertia to surface tension. We will keep the Prandtl number $Pr$ (a material property) and the Rayleigh number $Ra$ (the dimensionless temperature difference between the plates) fixed, at values allowing for considerable turbulence. Our main result will be the phase diagram in the parameter space ($We$, $\Lambda$), in which we identify the non-breakup and breakup regimes. At increasing $We$, we observe two distinct types of interface breakup. At small $\Lambda \ll 1$, the mechanism is well-described by the Kolmogorov-Hinze theory. However, at large $0.5<\Lambda \le 1$, the breakup is dominated by a balance between buoyancy and surface tension forces, leading to a periodic overturning-type breakup. 

The organization of this paper is as follows. The numerical methodology is introduced in Section \ref{meth}. Then in Section \ref{ver}, we validate our code by studying droplet fragmentation in thermal turbulence and favourably compare the results to the Kolmogorov-Hinze theory. The main results on the interface breakup in two-layer RB turbulence are presented in Section \ref{dis}, including a discussion of the first and second types of interface breakup in Section \ref{first} and Section \ref{second}, respectively, and the analysis of the critical Weber number for interface breakup in Section \ref{we}. We finalise and further discuss our findings in Section \ref{conc}.

\section{Methodology: Cahn-Hilliard approach coupled to finite difference scheme}\label{meth}

In this study, we performed the simulations in a two-dimensional (2D) rectangular domain with aspect ratio $\Gamma=2$ (width divided by height). Although 2D RB convection is different from 3D one, it still captures many essential features thereof \citep{2d}. The direct numerical simulations solver for the Navier-Stokes equations is a second-order finite-difference open source solver \citep{jcp96, cf15}, namely \href{https://github.com/PhysicsofFluids/AFiD}{AFiD}, which has been well validated and used to study various turbulent flows \citep{richard,zhu-tc,shan,qi,qi2}. To simulate multiphase turbulent flows, we combine AFiD with the phase-field method \citep{jaqcmin, ding, liu}, which has also been successfully applied to the interfacial \citep{liu2,zy, chen1,chen} and turbulent flows \citep{soldati1,soldati2}.

We consider two immiscible fluid layers of the same volume placed in the domain, named fluid $H$ for the heavier fluid initially at the bottom and fluid $L$ for the lighter fluid initially at the top. The mathematically sharp interface between two fluids is modeled by a diffuse interface with finite thickness, and can be represented by contours of the volume fraction $C$ of fluid $H$. The corresponding volume fraction of fluid $L$ is $1-C$. The evolution of $C$ is governed by the Cahn-Hilliard equations,
\begin{equation}
\frac {\partial C} {\partial t} + \nabla \cdot ({\bf u} C) = \frac{1}{Pe}\nabla^2 \psi,
\label{ch}
\end{equation} 
where $\bf u$ is the flow velocity, and $\psi= C^{3} - 1.5 C^{2}+ 0.5 C  -\hbox{Cn}^{2} \nabla^2 C$ is the chemical potential. The Cahn number $\hbox{Cn}=0.75h/D$, where $h$ is the mesh size and $D$ the domain height, and the P\'eclet number $Pe=0.9/\hbox{Cn}$ are set the same as in \cite{liu3,liu20}. To overcome the mass loss in the phase-field method, a correction method as proposed by \cite{shu} is used. This correction method resembles that used in \cite{soldati3} and exhibits good performance (see Section \ref{ver}).

The motion of the fluids is governed by the Navier-Stokes equation, heat transfer equation and continuity,

\begin{equation}
\rho \left(\frac {\partial {\bf u}} {\partial t} + {\bf u} \cdot \nabla {\bf u}\right)= - \nabla P + \sqrt {\frac{Pr}{Ra}} \nabla \cdot [\mu (\nabla {\bf u}+ \nabla {\bf u}^{T})] + {{\bf F}_{st}}+ \left(\rho\,\theta-\frac{\rho}{Fr}\right) {\bf j},
\label{ns}
\end{equation}

\begin{equation}
\frac{\partial {\theta}}{\partial t} + {\bf u} \cdot \nabla \theta =   \sqrt{\frac{1}{ Pr Ra }} \, \frac{1}{\rho C_p} \nabla \cdot (k \nabla \theta),
\label{t}
\end{equation}

\begin{equation}
\nabla \cdot {\bf u}= 0,
\label{con}
\end{equation}
{\color{black}here given in non-dimensionalized form. We have used the material properties of fluid $H$, the domain height $D$, the temperature difference $\Delta$ between plates, and the free-fall velocity $U=\sqrt{\alpha g D \Delta}$ to make these equations dimensionless, where $\alpha$ is the thermal expansion coefficient of fluid $H$ and $g$ the gravitational acceleration. Then we define} $\rho=C+\Lambda(1-C)$ as the {\color{black}dimensionless} density, $P$ the {\color{black}dimensionless} pressure, $\mu$ the {\color{black}dimensionless} dynamic viscosity, $C_p$ the {\color{black}dimensionless} specific heat capacity, $k$ the {\color{black}dimensionless} thermal conductivity, ${\bf F}_{st}=6\sqrt{2}\psi \nabla C / (\hbox{Cn\,We})$ the {\color{black}dimensionless} surface tension force, $\theta$ the dimensionless temperature, $\bf j$ the unit vector in vertical direction. {\color{black} The superscript $T$ stands for the transpose. Note that $\rho$, $\mu$, $C_p$ and $k$ in general vary in space.}

{\color{black}In thermal flows, the density also depends on the temperature, so the dimensional density is defined as $\hat{\rho}=\hat{\rho}_H(T)C+\hat{\rho}_L(T)(1-C)$, where the subscripts $H$ and $L$ represent fluid $H$ and fluid $L$, respectively, and $\hat{\rho}_i(T)=[1-\alpha(T-T_c)]\hat{\rho}_i(T_c)$ with $T_c$ being the temperature on the top cold plate. Then we rewrite the dimensional density as $\hat{\rho}=[\rho-\alpha(T-T_c)\rho]\hat{\rho}_H(T_c)$. Further considering the Oberbeck-Boussinesq approximation in the Navier-Stokes equation (\ref{ns}), we have the dimensionless density $\rho$ in the inertia term, $\rho \theta$, as the buoyancy term, and $\rho/Fr$ as the gravity term, which cannot be ignored as in the single phase simulation, due to the different densities of the fluids. Furthermore, we only consider the case without phase transition. 

The properties of fluid $H$ and fluid $L$ are set as follows:} Their density ratio is $\Lambda=\rho_L/\rho_H \le 1$. Except for the density $\rho$, all other properties of the two fluids are the same. The other dimensionless parameters are $We=\rho_H U^2 D/\sigma$, $Ra=\alpha g D^3 \Delta/(\nu \kappa)$, $Pr=\nu/\kappa$, the Froude number $Fr=U^2/(gD)$ 
(the ratio of inertia to gravity). Here $\sigma$ is the surface tension coefficient, $\nu=\mu/\rho$ the kinematic viscosity, and $\kappa=k/(\rho C_p)$ the thermal diffusivity. We fix $Ra=10^8$, $Pr=1$, $Fr=1$ and $\Gamma=2$ (these values chosen to both 
ensure flows in the turbulent regime and simplify the simulations), and only vary $We$ from $5$ to $5000$ and $\Lambda$ from $0.001$ (e.g. air-water system) to $1$ (e.g. oil-water system). 

The boundary conditions at the top and bottom plates are set as $\partial C/\partial {\bf j}=0$, {\color{black}${\bf j}\cdot\nabla \psi=0$}, no-slip velocities and fixed temperature $\theta=0$ (top) and $1$ (bottom). Periodic conditions are used in the horizontal direction. All the simulations begin with the same initial velocity and temperature fields, which originates from a well developed turbulent flow at $We=5$ and $\Lambda=1$.  Uniform grids with $1000 \times 500$ gridpoints are used, which are sufficient for $Ra=10^8$ and $Pr=1$, consistent with the grid resolution checks in \cite{zhou}. {\color{black} The details of the discretizations can be found in \cite{ding,jcp96,jcp14}.}

\section{Droplet fragmentation in turbulent flow}
\label{ver}
We have verified our code against the existing theory from the literature. In this section RB convection with drops are simulated. Initially, the temperature field has a linear profile, the velocity field is set to $0$, and a big drop of fluid $H$ with a diameter of $0.5D$ is placed at the center of the domain. The plates are superhydrophobic for fluid $H$, i.e. $C=0$ is used on both plates in the verification cases. We set $\Lambda=1$ 
and $We$ from $2000$ to $16000$, and use uniform grids with $2000 \times 1000$ gridpoints. Note that the value of $We$ is large because it is a system Weber number, and the local Weber number of drops calculated from simulations is of the order of $1$.
 
\begin{figure}
\centering
\includegraphics[width=0.6\linewidth]{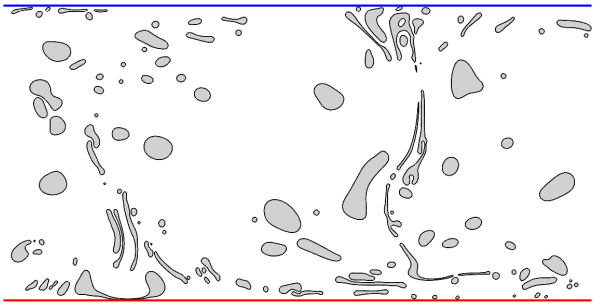}
\caption{\label{fig1}   Snapshot with the advecting drops in Rayleigh-B\'enard convection at density ratio $\Lambda = 1$ and the system Weber number $We=16000$. Drops are in gray, and the red and blue lines denote the plates with non-dimensional temperature $\theta=1$ and $0$, respectively. The corresponding movie is shown as Supplementary Material.}
\end{figure}

\begin{figure}
\centering
\includegraphics[width=1\linewidth]{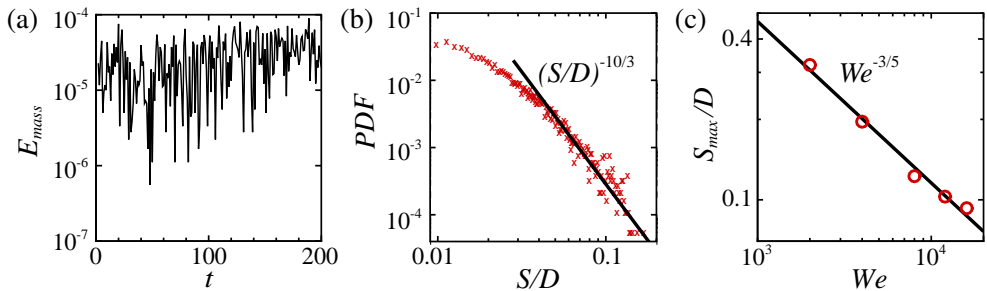}
\caption{\label{fig2}   (a) Temporal evolution of Mass error $E_{mass}$ and (b) probability density function (PDF) of the drop size $S/D$ at $We=16000$, where $D$ is the domain height and $S=2\sqrt{A/\pi}$ with $A$ being the drop area. (c) Maximal drop size $S_{max}/D$ as function of $We$, where $S_{max}$ is measured in the same way as in \cite{hinze}, that is the diameter of the equivalent drop occupying $95\%$ of the total dispersed area.}
\end{figure}
 
From the snapshot with the advecting drops in figure \ref{fig1}, we observe the large scale circulation, which is well known from single phase convection  \citep{lsc81, xi04jfm, zhu, xie2}. Figure \ref{fig2}(b) displays the distribution of the drop size $S$, which follows the scaling law $(S/D)^{-10/3}$ \citep{pdf, pdfjfm} valid for large drops. This scaling is consistent with experimental and other numerical results where the breakup of waves \citep{nature02, pop16} or of a big drop \citep{LB19JFM} was studied. Moreover, the maximal size of drops $S_{max}$ (see figure \ref{fig2}c) well agrees with the Kolmogorov-Hinze scaling law $S_{max}/D \sim We^{-3/5}$, which originates from \cite{hinze}, 
\begin{equation}
S_{Hinze} \sim \left(\frac{\sigma}{\rho}\right)^\frac{3}{5}\epsilon^{-\frac{2}{5}},
\label{Kolmogorov-Hinze}
\end{equation} 
where $S_{Hinze}$ is the Hinze length scale and $\epsilon$ the energy dissipation rate of the turbulent flow. {\color{black} In the Kolmogorov-Hinze theory, one assumption is that the local Weber number defined by the size and velocity of the drops adjusts such that it is $We_{local} \sim O(1)$. Indeed, in our simulations the local drop size adjusts correspondingly, so this assumption is fulfilled. The second assumption is that the flow exhibits inertial subrange scaling in the region of wave lengths comparable to the size of the largest drops. The spatial location where this assumption holds in RB convection is in the bulk region of convection \citep{rev2}. Therefore, the Kolmogorov-Hinze theory can indeed also be reasonably applied to 2D RB convection.} At the same time the results show that the code and the approach give consistent results. Besides, figure \ref{fig2}(a) shows the temporal evolution of mass error $E_{mass}=|M_t-M_0|/M_0$, where $M_t$ is the mass of fluid $H$ at time $t$ and and $M_0$ the initial mass. The results indicate good mass conservation.

\section{Interfacial breakup in two-layer turbulent thermal convection}\label{dis}
 
\begin{figure}
\centering
\includegraphics[width=\linewidth]{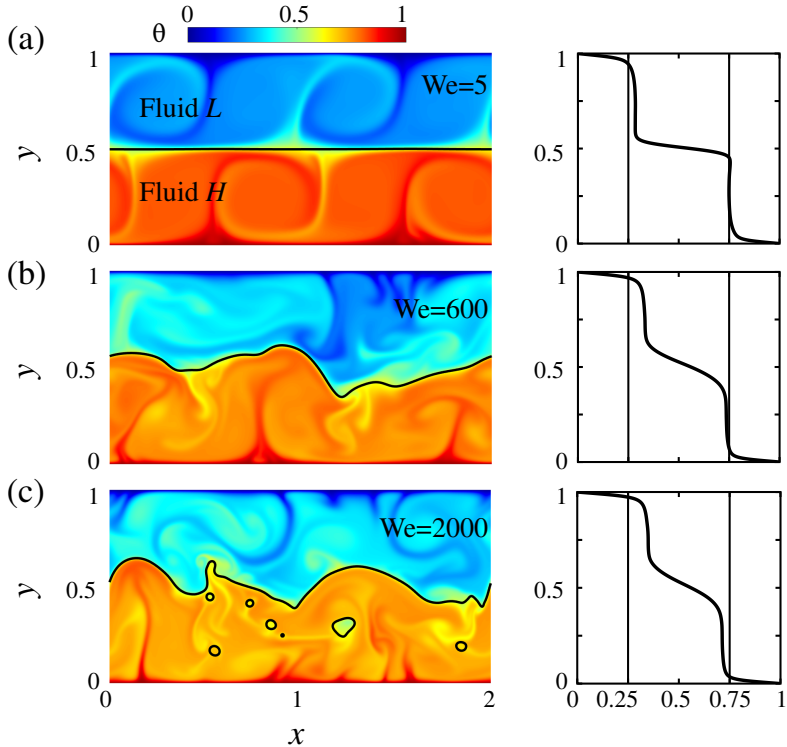}
\caption{\label{fig3}  First type of interface breakup occurring for small $\Lambda \ll 1$: Temperature field and average temperature profile of two-layer Rayleigh-B\'enard convection at $\Lambda = 0.3$ for (a) $We = 5$, (b) $We = 600$ and (c) $We = 2000$. The corresponding movies are shown as Supplementary Material.}
\end{figure}
 
\begin{figure}
\centering
\includegraphics[width=0.9\linewidth]{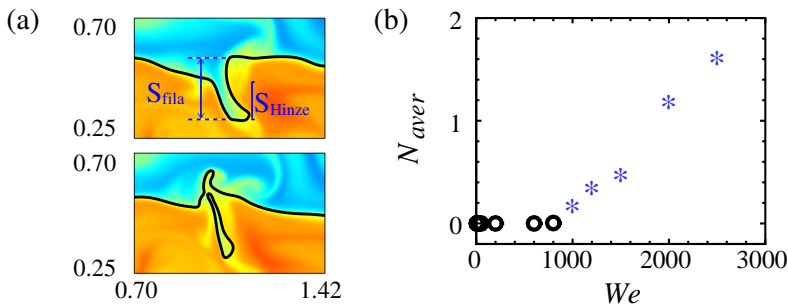}
\caption{\label{fig4}  (a) Detachment process of a drop at $We=2000$. (b) Time-averaged number of the drops of fluid $L$ emerged in fluid $H$ for various $We$, where the empty circles denote the non-breakup regime and stars the breakup regime. }
\end{figure}

In two-layer RB convection with an initially smooth interface between the two fluids \citep{xie}, the densities of the fluids depend on both $\Lambda$ and the local temperature. At small $\Lambda$, for example, the air-water system with $\Lambda=0.001$, fluid $H$ (water) is always heavier than fluid $L$ (air) even if fluid $H$, as the bottom layer, is hotter. In contrast, at large $\Lambda$, e.g., an oil-water system with $\Lambda \approx 1$, fluid $H$ (water) is hotter than fluid $L$ (oil), so it can be lighter. So depending on the value of $\Lambda$ two distinct types of flow phenomena emerge.

\subsection{First type of interface breakup occurring for small $\Lambda \ll 1$}
\label{first}

At small $\Lambda$, fluid $H$ forms the bottom layer and fluid $L$ the top one and large scale circulations are observed in each layer between the interface and the respective plate, as seen in figure \ref{fig3}(a). With increasing $We$, i.e. decreasing effects of surface tension compared to inertia, the interface becomes more unstable. At low $We$ (figure \ref{fig3}a), the interface only slightly deforms because the surface tension is large enough to resist the inertia, such that the convection rolls are well-ordered. The temperature profile is similar to that obtained from two-layer RB convection experiments \citep{nature99}. As $We$ increases (figure \ref{fig3}b), the interface undulates due to the plumes. Each crest and trough on the interface is caused by a rising, or respectively settling plume in the heavier fluid $H$. In this situation, inertia is resisted by gravity together with surface tension. When $We$ keeps increasing (figure \ref{fig3}c), the interface eventually breaks up and drops detach from the interface between two layers.

This ``first type of interface breakup" (as we call it) occurs at small $\Lambda$. The process of the breakup begins from a settling plume in fluid $H$ thanks to which the interface is pulled downwards, leading to a filament (or trough) on the interface (see figure \ref{fig4}a). If the filament length $S_{fila}$ (defined in figure \ref{fig4}a) grows larger than the Hinze length scale $S_{Hinze}$, the filament will pinch off from the interface. Within the Kolmogrov-Hinze theory \citep{hinze}, the Hinze length scale $S_{Hinze}$ in (\ref{Kolmogorov-Hinze}) is determined by the energy dissipation rate $\epsilon$ of the turbulent flow. As seen from figure \ref{fig3}(c), more drops of fluid $L$ exist in fluid $H$ than of fluid $H$ in fluid $L$. This finding resembles the breakup of the ocean waves \citep{nature02}, leading to more bubbles in water than drops in air. The reason is that $S_{Hinze}$ is smaller in fluid $H$ than in fluid $L$ as $\rho_H>\rho_L$.

\subsection{Second type of interface breakup occurring for large $0.5<\Lambda \le 1$}
\label{second}

 \begin{figure}
\centering
 \includegraphics[width=1\linewidth]{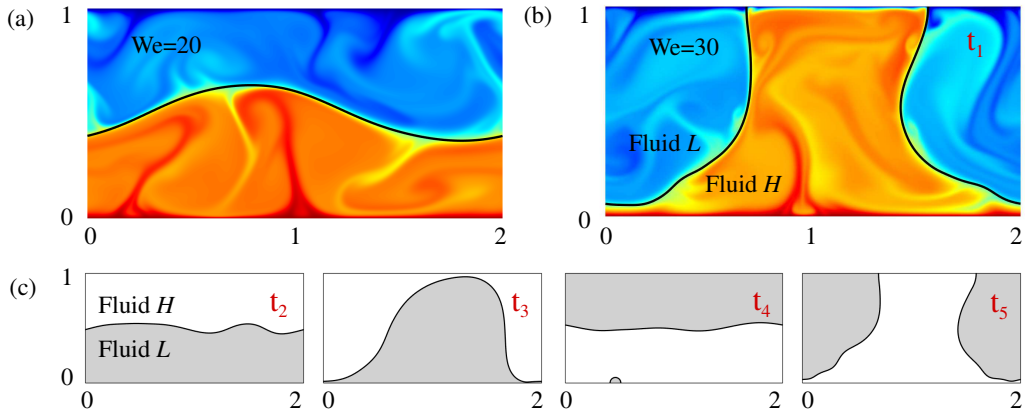}
 \caption{\label{fig5}   Second type of interface breakup occurring for large $0.5<\Lambda \le 1$: Snapshots at $\Lambda = 0.8$ for two different $We$. (a) Wavy interface for $We = 20$. (b) Breakup and (c) overturning of interface for $We = 30$ at different times $t_1=617$, $t_2=640$, $t_3=661$, $t_4=730$ and $t_5=803$. $t_i$ are also marked in figure \ref{fig6}. The color map is the same as in figure \ref{fig3}. The corresponding movies are shown as Supplementary Material. }
 \end{figure}

 \begin{figure}
\centering
 \includegraphics[width=\linewidth]{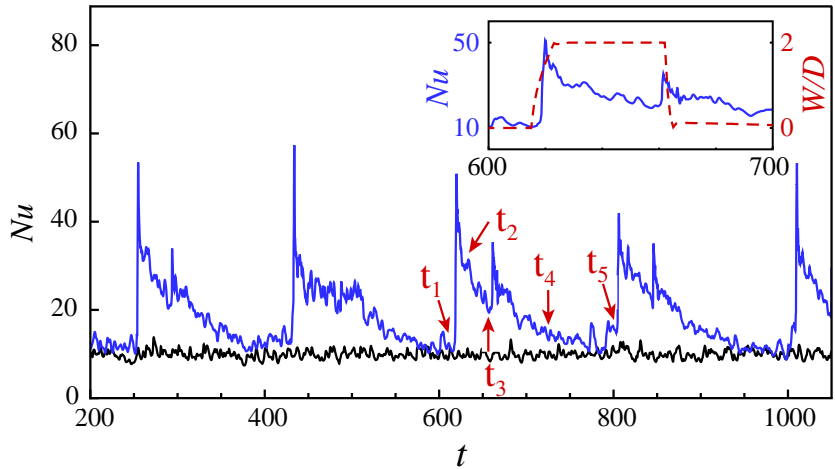}
 \caption{\label{fig6}   Temporal evolution of the Nusselt number $Nu$ at the bottom plate for $We=20$ (black) and $We=30$ (blue). The inset shows a zoom of the temporal evolution of $Nu$ for $We=30$ and the corresponding wetted length function $W/D$ of fluid $H$ at the top plate.}
 \end{figure}

We now come to the large $\Lambda \approx 1$ case: Since fluid $H$ carries hotter fluid than fluid $L$, due to thermal expansion it can become lighter than fluid $L$, inverting the original density contrast at equal temperature. In this situation, buoyancy drives fluid $H$ upwards and fluid $L$ downwards. This leads to wave crests and troughs, as shown in figure \ref{fig5}. If $We$ is low (figure \ref{fig5}a), the surface tension can maintain a stable interface, though it is wobbling. However, if $We$ increases (figure \ref{fig5}b), the wobbling wave on the interface can amplify more and more until it finally touches the upper and/or lower plate and breaks up. We call this type of interface breakup the ``second type of breakup".

For this second type of interface breakup, the breakup process is strikingly different from the first type. A periodic overturning is observed both in the fluid dynamics and in the heat transfer (see figure \ref{fig5}c and figure \ref{fig6}): After interface breakup at $t_1$, fluid $H$ initially at the bottom gradually rises above fluid $L$ and finally contacts the upper cold plate. The increased wetted area of the hotter fluid on the upper cold plate causes a strong enhancement of the Nusselt number $Nu$, in the shown case $5$ times of $Nu$ without breakup, as shown in figure \ref{fig6}. Then, fluid $H$ on the top gets cooler and thus heavier, while fluid $L$ at the bottom warmer and thus lighter. Once again, the breakup occurs after $t_3$ and the fluids swap their positions (see figure \ref{fig5}c). Since fluid $L$ is lighter than fluid $H$ at the same temperature, it is easier to rise and touch the cold plate. Thus, with fluid $L$ as the bottom layer, the temperature of the bottom layer during the breakup is lower than that when fluid $H$ is the bottom layer. This is also reflected in the heat transfer. The peak of $Nu$ after $t_3$ is smaller, only 3 times of $Nu$ without breakup, and the preparation time for breakup from $t_2$ to $t_3$ is shorter than that from $t_4$ to $t_5$.
\\

\subsection{
Critical Weber number for interface breakup}
\label{we}

 \begin{figure}
\centering
 \includegraphics[width=\linewidth]{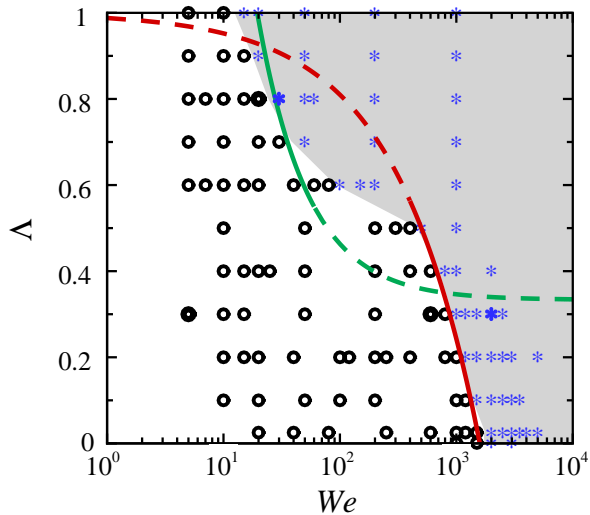}
 \caption{\label{fig7}   Phase diagram in the $We-\Lambda$ parameter space. Empty circles denote the non-breakup regime and stars the interface breakup regime. Symbols with boldface are the cases shown in figure \ref{fig3} and \ref{fig5}. The gray shadow is a guide to the eye. The red and green lines denote the criteria, (\ref{eq3}) with prefactor $1590$ and (\ref{eq6}) with prefactor $13.3$, for the first and second type of interface breakup, respectively. The solid parts of the lines, where the theory is supposed to hold, indeed nicely agree with the numerical results.}
 \end{figure}
 
The full phase diagram in the parameter space ($We, \Lambda$) -- revealing when and what regime occurs -- is plotted in figure \ref{fig7}. When $We$ is larger than a certain critical value $We_c$, which depends on $\Lambda$, the interface breaks up. It is noteworthy that the transition between the non-breakup and breakup regimes show two distinct trends, which correspond to the two above identified types of interface breakup, respectively. The natural question that arises here is then: What sets the critical value $We_c$ at given $\Lambda$?

In the first type of interface breakup (small $\Lambda \ll 1$), detaching drops are generated from the initial interface when the filament length $S_{fila}$ is of the order of the Hinze length scale $S_{Hinze}$. $S_{fila}$ is estimated by analyzing the force balance: the sum of gravity and surface tension force counteracts the inertial force,
\begin{equation}
\delta \rho \, gS_{fila} + \frac{\sigma}{S_{fila}} \sim \rho_H U_H^2, 
\label{eq1}
\end{equation}
where $U_H=\sqrt{\alpha g (D/2) (\Delta/2)}$ and $\delta \rho$ is the density difference from the bottom (fluid $H$) to the top (fluid $L$) of the filament. We define $\delta \rho=\rho_H({T_H}) - \rho_L({T_L})$, where $T_i$ is the temperature of fluid $i$.
We further found that the value of gravity is one order of magnitude greater than surface tension based on the data of cases near the transition region in the first type. This indicates that the generation of the filament is dominated by gravity and inertia. Therefore we neglect the term $\sigma/S_{fila}$ in (\ref{eq1}). Note however that the surface tension force still plays an important role to determine $S_{Hinze}$ in (\ref{Kolmogorov-Hinze}). Combining (\ref{eq1}), (\ref{Kolmogorov-Hinze}) and the exact relation $\epsilon=\nu^3/D^4 \, (Nu-1) Ra Pr^{-2}$ \citep{pra1990, rev1}, the dimensionless form of $S_{fila} \sim S_{Hinze}$ yields
\begin{equation}
\left[\left(\frac{1}{Fr}-\theta_H\right)-\Lambda\left(\frac{1}{Fr}-\theta_L \right)\right]^{-1} \sim We_c^{-\frac{3}{5}}\left(\frac{Nu-1}{\sqrt{RaPr}}\right)^{-\frac{2}{5}},
\label{eq2}
\end{equation}
with the non-dimensional temperatures $\theta_i=(T_i-T_c)/\Delta$ with $i$ being $H$ and $L$. $\theta_H$ and $\theta_L$ are both taken as $0.5$ given that the filament is generated near the interface, where the temperature is $0.5$. To further simplify (\ref{eq2}), $Nu$ is regarded as constant because the simulation data show that $Nu$ varies only within $15\%$ in the non-breakup regime. Given that the $Nu$, $Ra$, $Pr$ and $Fr$ are all constant, the criteria for the first type of interface breakup simplifies to 
\begin{equation}
We_c \sim \left(1-\Lambda \right)^{\frac{5}{3}}.
\label{eq3}
\end{equation}

In the second type of interface breakup (large $\Lambda \approx 1$), the hot fluid $H$ is lighter than the cold fluid $L$, so the buoyancy caused by the unstable temperature stratification can overcome the surface tension, leading to waves on the interface. The buoyancy acting on the wave originates from the density difference between the fluid above and below the wobbling interface (see figure \ref{fig5}). The balance is described by
\begin{equation}
 [\rho_L({T_L}) - \rho_H({T_H})]gD \sim \frac{\sigma}{D},
\label{eq4}
\end{equation}
where $T_i$ is the average temperature in the bulk of fluid $i$, and the interface deformation is of the order of $D$ because the breakup occurs when the wave amplitude is larger than half of the plate distance $D$, thus that the interface touches the plates (see figure \ref{fig5}b). The dimensionless form of (\ref{eq4}) reads
\begin{equation}
 \Lambda \left(\frac{1}{Fr}-\theta_L\right)-\left(\frac{1}{Fr}-\theta_H\right) \sim \frac{1}{We_c}.
\label{eq5}
\end{equation}
From the temperature profile in figure \ref{fig3}, we estimate $\theta_H=0.75$ and $\theta_L=0.25$. Then (\ref{eq5}) simplifies to
\begin{equation}
We_c \sim \left(\Lambda-\frac{1}{3}\right)^{-1}.
\label{eq6}
\end{equation}
Figure \ref{fig7} shows that (\ref{eq3}) and (\ref{eq6}) indeed well describe the scaling relations of transitions between the non-breakup and breakup regimes. 

{\color{black}In the breakup regimes at large $\Lambda \sim 1$, as $We$ increases, the periodically overturning of fluid layers gradually becomes chaotic with more and more drops generated from the breakup of the interface. Eventually, the flow pattern is determined by the advection of the small drops (figure \ref{fig1}). There is no clear boundary in the parameter space for this transition and it happens over a considerable range in parameter space. Therefore, in this study, we only focus on when and how the breakup occurs.}

\section{Conclusions}\label{conc}

In summary, we have numerically shown two distinct types of interface breakup in two-layer RB convection, which result from different dominant forces. At small $\Lambda \ll 1$, a filament is generated on the interface due to the competition of inertial force and buoyancy. The interface breaks up in the form of filament detachment when the local filament thicknesses exceed the Hinze length scale. At large $\Lambda$ with roughly $0.5 < \Lambda \le 1$, the periodic overturning-type breakup is caused by buoyancy overwhelming surface tension. From the force balance arguments above, we derive the breakup criteria for two types, respectively. Our approaches show good agreements with the numerical results.

{\color{black}The threshold of regimes in this work is derived from a force balance argument, which is not limited to 2D. For 3D, of course, some expressions need to be modified, for example, surface tension force from $F_{st}=\sigma/R$ in 2D to $F_{st}=\sigma(1/R_1+1/R_2)$ in 3D. Besides, we also note that previous studies also showed that 2D and 3D Rayleigh-B\'enard convection have similar features for $Pr\ge1$, see \cite{2d}. Therefore, we expect our results from 2D flow could be directly extended to 3D flow.}

Our findings clearly demonstrate that interface breakup in multilayer thermally driven turbulence cannot be solely described by the Kolmogorov-Hinze theory, which is only applicable when the lower layer is much denser than the upper layer or when thermal expansion effects do not play a role. Interestingly, when the lower layer is less dense than the upper layer, the system is unstably stratified, leading to the new breakup type described in this paper where buoyancy and surface tension are the dominant forces. It would be interesting to test our predictions of (\ref{eq2}) and (\ref{eq5}) for the transitions between the different regimes in a larger range of the control parameters $Ra$, $Pr$, and $Fr$, which were kept fixed here. More generally, our findings emphasize the role of buoyancy in interfacial breakup. Clearly, buoyancy will also play a prominent role in the interfacial breakup in other turbulent flows, such as B\'enard-Marangoni convection, and horizontal and vertical convection. All these phenomena add to the richness of physicochemical hydrodynamics of droplets far from equilibrium \citep{nrp2020}.

\section*{Acknowledgments}
The work was financially supported by ERC-Advanced Grant under the project no.~$740479$. We acknowledge PRACE for awarding us access to MareNostrum in Spain at the Barcelona Computing Center (BSC) under the project $2018194742$. This work was also carried out on the national e-infrastructure of SURFsara, a subsidiary of SURF cooperation, the collaborative ICT organization for Dutch education and research.

\section*{Declaration of interests}
The  authors  report  no  conflict  of  interest.

\section*{Supplementary movies}
Supplementary  movies  are  available at \href{https:}{URL}

\bibliographystyle{jfm}

\begin{thebibliography}{60}
\expandafter\ifx\csname natexlab\endcsname\relax\def\natexlab#1{#1}\fi
\def\au#1{#1} \def\ed#1{#1} \def\yr#1{#1}\def\at#1{#1}\def\jt#1{\textit{#1}}
  \def\bt#1{#1}\def\bvol#1{\textbf{#1}} \def\vol#1{#1} \def\pg#1{#1}
  \def\publ#1{#1}\def\arxiv#1{#1}\def\org#1{#1}\def\st#1{\textit{#1}}

\bibitem[Ahlers {\em et~al.\/}(2009)Ahlers, Grossmann \& Lohse]{rev1}
{\sc \au{Ahlers, G.}, \au{Grossmann, S.} \& \au{Lohse, D.}} \yr{2009}  \at{Heat
  transfer and large scale dynamics in turbulent {Rayleigh-B{\'{e}}nard}
  convection}.  \jt{Rev. Mod. Phys.}  \bvol{81},  \pg{503}.

\bibitem[Blass {\em et~al.\/}(2020)Blass, Zhu, Verzicco, Lohse \&
  Stevens]{shan}
{\sc \au{Blass, A.}, \au{Zhu, X.}, \au{Verzicco, R.}, \au{Lohse, D.} \&
  \au{Stevens, R.~J.A.M.}} \yr{2020}  \at{Flow organization and heat transfer
  in turbulent wall sheared thermal convection}.  \jt{J. Fluid Mech.}
  \bvol{897},  \pg{A22}.

\bibitem[Bourouiba(2020)]{covid}
{\sc \au{Bourouiba, L.}} \yr{2020}  \at{Turbulent gas clouds and respiratory
  pathogen emissions: Potential implications for reducing transmission of
  covid-19}.  \jt{JAMA} .

\bibitem[Bourouiba {\em et~al.\/}(2014)Bourouiba, Dehandschoewercker \&
  Bush]{sneezing}
{\sc \au{Bourouiba, L.}, \au{Dehandschoewercker, E.} \& \au{Bush, J. W.~M.}}
  \yr{2014}  \at{Violent expiratory events: on coughing and sneezing}.  \jt{J.
  Fluid Mech.}  \bvol{745},  \pg{537--563}.

\bibitem[Busse \& Petry(2009)]{low3}
{\sc \au{Busse, F.~H.} \& \au{Petry, M.}} \yr{2009}  \at{Homologous onset of
  double layer convection}.  \jt{Phys. Rev. E}  \bvol{80},  \pg{046316}.

\bibitem[Chen {\em et~al.\/}(2020)Chen, Liu, Gao \& Ding]{chen}
{\sc \au{Chen, H.}, \au{Liu, H.-R.}, \au{Gao, P.} \& \au{Ding, H.}} \yr{2020}
  \at{Submersion of impacting spheres at low {Bond} and {Weber} numbers owing
  to a confined pool}.  \jt{J. Fluid Mech.}  \bvol{884},  \pg{A13}.

\bibitem[Chen {\em et~al.\/}(2018)Chen, Liu, Lu \& Ding]{chen1}
{\sc \au{Chen, H.}, \au{Liu, H.-R.}, \au{Lu, X.-Y.} \& \au{Ding, H.}} \yr{2018}
   \at{Entrapping an impacting particle at a liquid-gas interface}.  \jt{J.
  Fluid Mech.}  \bvol{841},  \pg{1073--1084}.

\bibitem[Chill{\`{a}} \& Schumacher(2012)]{chilla}
{\sc \au{Chill{\`{a}}, F.} \& \au{Schumacher, J.}} \yr{2012}  \at{New
  perspectives in turbulent {Rayleigh-B{\'{e}}nard} convection}.  \jt{Eur.
  Phys. J. E}  \bvol{35},  \pg{58}.

\bibitem[Chong {\em et~al.\/}(2020)Chong, Ng, Hori, Yang, Verzicco \&
  Lohse]{covid-steven}
{\sc \au{Chong, K.~L.}, \au{Ng, C.~S.}, \au{Hori, N.}, \au{Yang, R.},
  \au{Verzicco, R.} \& \au{Lohse, D.}} \yr{2020}  \at{Extended lifetime of
  respiratory droplets in a turbulent vapour puff and its implications on
  airborne disease transmission}.  \jt{arXiv preprint arXiv:2008.01841} .

\bibitem[Davaille(1999)]{nature99}
{\sc \au{Davaille, A.}} \yr{1999}  \at{Simultaneous generation of hotspots and
  superswells by convection in a heterogeneous planetary mantle}.  \jt{Nature}
  \bvol{402},  \pg{756--760}.

\bibitem[Deane \& Stokes(2002)]{nature02}
{\sc \au{Deane, G.~B.} \& \au{Stokes, M.~D.}} \yr{2002}  \at{Scale dependence
  of bubble creation mechanisms in breaking waves}.  \jt{Nature}  \bvol{418},
  \pg{839--844}.

\bibitem[Deike {\em et~al.\/}(2016)Deike, Melville \& Popinet]{pop16}
{\sc \au{Deike, L.}, \au{Melville, W.~K.} \& \au{Popinet, S.}} \yr{2016}
  \at{Air entrainment and bubble statistics in breaking waves}.  \jt{J. Fluid
  Mech.}  \bvol{801},  \pg{91--129}.

\bibitem[Ding {\em et~al.\/}(2007)Ding, Spelt \& Shu]{ding}
{\sc \au{Ding, H.}, \au{Spelt, P. D.~M.} \& \au{Shu, C.}} \yr{2007}
  \at{Diffuse interface model for incompressible two-phase flows with large
  density ratios}.  \jt{J. Comput. Phys.}  \bvol{226},  \pg{2078--2095}.

\bibitem[Diwakar {\em et~al.\/}(2014)Diwakar, Tiwari, Das \&
  Sundararajan]{low-ra}
{\sc \au{Diwakar, S.~V.}, \au{Tiwari, S.}, \au{Das, S.~K.} \& \au{Sundararajan,
  T.}} \yr{2014}  \at{Stability and resonant wave interactions of confined
  two-layer {Rayleigh–B{\'{e}}nard} systems}.  \jt{J. Fluid Mech.}
  \bvol{754},  \pg{415--455}.

\bibitem[Garrett {\em et~al.\/}(2000)Garrett, Li \& Farmer]{pdf}
{\sc \au{Garrett, C.}, \au{Li, M.} \& \au{Farmer, D.}} \yr{2000}  \at{The
  connection between bubble size spectra and energy dissipation rates in the
  upper ocean}.  \jt{J. Phys. Oceanogr.}  \bvol{30}~(9),  \pg{2163--2171}.

\bibitem[Hesketh {\em et~al.\/}(1991)Hesketh, Etchells \& Russell]{pipe}
{\sc \au{Hesketh, R.~P.}, \au{Etchells, A.~W.} \& \au{Russell, T. W.~F.}}
  \yr{1991}  \at{Bubble breakage in pipeline flow}.  \jt{Chem. Eng. Sci.}
  \bvol{46}~(1),  \pg{1--9}.

\bibitem[Hinze(1955)]{hinze}
{\sc \au{Hinze, J.~O.}} \yr{1955}  \at{Fundamentals of the hydrodynamic
  mechanism of splitting in dispersion processes}.  \jt{AIChE J.}
  \bvol{1}~(3),  \pg{289--295}.

\bibitem[Jacqmin(1999)]{jaqcmin}
{\sc \au{Jacqmin, D.}} \yr{1999}  \at{Calculation of two-phase
  {Navier–Stokes} flows using {Phase-Field} modeling}.  \jt{J. Comput. Phys.}
   \bvol{155},  \pg{96--127}.

\bibitem[Josserand \& Zaleski(2003)]{zaleski}
{\sc \au{Josserand, C.} \& \au{Zaleski, S.}} \yr{2003}  \at{Droplet splashing
  on a thin liquid film}.  \jt{Phys. Fluids}  \bvol{15},  \pg{1650}.

\bibitem[Kolmogorov(1949)]{kol}
{\sc \au{Kolmogorov, A.~N.}} \yr{1949}  \at{On the disintegration of drops in a
  turbulent flow}.  \jt{Dokl. Akad. Navk. SSSR}  \bvol{66},  \pg{825--828}.

\bibitem[Krishnamurti \& Howard(1981)]{lsc81}
{\sc \au{Krishnamurti, R.} \& \au{Howard, L.~N.}} \yr{1981}  \at{Large-scale
  flow generation in turbulent convection}.  \jt{Proc. Natl. Acad. Sci. USA}
  \bvol{78}~(4),  \pg{1981--1985}.

\bibitem[Li {\em et~al.\/}(2020)Li, Liu \& Ding]{liu20}
{\sc \au{Li, H.-L.}, \au{Liu, H.-R.} \& \au{Ding, H.}} \yr{2020}  \at{A fully
  3d simulation of fluid-structure interaction with dynamic wetting and contact
  angle hysteresis}.  \jt{J. Comput. Phys.}  \bvol{420},  \pg{109709}.

\bibitem[Liu \& Ding(2015)]{liu}
{\sc \au{Liu, H.-R.} \& \au{Ding, H.}} \yr{2015}  \at{A diffuse-interface
  immersed-boundary method for two-dimensional simulation of flows with moving
  contact lines on curved substrates}.  \jt{J. Comput. Phys.}  \bvol{294},
  \pg{484--502}.

\bibitem[Liu {\em et~al.\/}(2017)Liu, Gao \& Ding]{liu3}
{\sc \au{Liu, H.-R.}, \au{Gao, P.} \& \au{Ding, H.}} \yr{2017}
  \at{Fluid-structure interaction involving dynamic wetting: {2D} modeling and
  simulations}.  \jt{J. Comput. Phys.}  \bvol{348},  \pg{45--65}.

\bibitem[Liu {\em et~al.\/}(2018)Liu, Zhang, Gao, Lu \& Ding]{liu2}
{\sc \au{Liu, H.-R.}, \au{Zhang, C.-Y.}, \au{Gao, P.}, \au{Lu, X.-Y.} \&
  \au{Ding, H.}} \yr{2018}  \at{On the maximal spreading of impacting compound
  drops}.  \jt{J. Fluid Mech}  \bvol{854},  \pg{R6}.

\bibitem[Lohse \& Xia(2010)]{rev2}
{\sc \au{Lohse, D.} \& \au{Xia, K.-Q.}} \yr{2010}  \at{Small-scale properties
  of turbulent {Rayleigh-B{\'{e}}nard} convection}.  \jt{Annu. Rev. Fluid
  Mech.}  \bvol{42},  \pg{335}.

\bibitem[Lohse \& Zhang(2020)]{nrp2020}
{\sc \au{Lohse, D.} \& \au{Zhang, X.}} \yr{2020}  \at{Physicochemical
  hydrodynamics of droplets out of equilibrium}.  \jt{Nat. Rev. Phys.}
  \bvol{2},  \pg{426--443}.

\bibitem[M.~S.~Dodd(2014)]{jcp14}
{\sc \au{M.~S.~Dodd, A.~Ferrante}} \yr{2014}  \at{A fast pressure-correction
  method for incompressible two-fluid flows}.  \jt{J. Comput. Phys.}
  \bvol{273},  \pg{416--434}.

\bibitem[Mart{\'{i}}nez-Baz{\'{a}}n {\em
  et~al.\/}(1999)Mart{\'{i}}nez-Baz{\'{a}}n, Monta{\~{n}}{\'{e}}s \&
  Lasheras]{hit99exp}
{\sc \au{Mart{\'{i}}nez-Baz{\'{a}}n, C.}, \au{Monta{\~{n}}{\'{e}}s, J.~L.} \&
  \au{Lasheras, J.~C.}} \yr{1999}  \at{On the breakup of an air bubble injected
  into a fully developed turbulent flow. {Part 2. Size PDF} of the resulting
  daughter bubbles}.  \jt{J. Fluid Mech.}  \bvol{401},  \pg{183--207}.

\bibitem[MuKolmogorov-Hinzeerjee {\em et~al.\/}(2019)MuKolmogorov-Hinzeerjee, Safdari, Shardt,
  Kenjere{\v{s}} \& {Van den Akker}]{LB19JFM}
{\sc \au{MuKolmogorov-Hinzeerjee, S.}, \au{Safdari, A.}, \au{Shardt, O.}, \au{Kenjere{\v{s}},
  S.} \& \au{{Van den Akker}, H. E.~A.}} \yr{2019}  \at{Droplet-turbulence
  interactions and quasi-equilibrium dynamics in turbulent emulsions}.  \jt{J.
  Fluid Mech.}  \bvol{878},  \pg{221--276}.

\bibitem[Nataf {\em et~al.\/}(1988)Nataf, Moreno \& Cardin]{low1}
{\sc \au{Nataf, H.~C.}, \au{Moreno, S.} \& \au{Cardin, P.}} \yr{1988}  \at{What
  is responsible for thermal coupling in layered convection?}  \jt{J. Phys.
  (Paris)}  \bvol{49},  \pg{1707--1714}.

\bibitem[Perlekar {\em et~al.\/}(2012)Perlekar, Biferale, Sbragaglia,
  Srivastava \& Toschi]{bt}
{\sc \au{Perlekar, P.}, \au{Biferale, L.}, \au{Sbragaglia, M.}, \au{Srivastava,
  S.} \& \au{Toschi, F.}} \yr{2012}  \at{Droplet size distribution in
  homogeneous isotropic turbulence}.  \jt{Phys. Fluids}  \bvol{24},
  \pg{065101}.

\bibitem[Prakash \& Koster(1994)]{low2}
{\sc \au{Prakash, A.} \& \au{Koster, J.~N.}} \yr{1994}  \at{Convection in
  multiple layers of immiscible liquids in a shallow cavity {I}. steady natural
  convection}.  \jt{Intl J. Multiphase Flow}  \bvol{20}~(2),  \pg{383--396}.

\bibitem[Roccon {\em et~al.\/}(2019)Roccon, Zonta \& Soldati]{soldati2}
{\sc \au{Roccon, A.}, \au{Zonta, F.} \& \au{Soldati, A.}} \yr{2019}
  \at{Turbulent drag reduction by compliant lubricating layer}.  \jt{J. Fluid
  Mech.}  \bvol{863},  \pg{R1}.

\bibitem[Rosti {\em et~al.\/}(2019)Rosti, Ge, Jain, Dodd \& Brandt]{luka19jfm}
{\sc \au{Rosti, M.~E.}, \au{Ge, Z.}, \au{Jain, S.~S.}, \au{Dodd, M.~S.} \&
  \au{Brandt, L.}} \yr{2019}  \at{Droplets in homogeneous shear turbulence}.
  \jt{J. Fluid Mech.}  \bvol{876},  \pg{962--984}.

\bibitem[Shraiman \& Siggia(1990)]{pra1990}
{\sc \au{Shraiman, B.~I.} \& \au{Siggia, E.~D.}} \yr{1990}  \at{Heat transport
  in high rayleigh number convection}.  \jt{Phys. Rev. A}  \bvol{42},
  \pg{3650--3653}.

\bibitem[Soligo {\em et~al.\/}(2019{\natexlab{{\em a\/}}})Soligo, Roccon \&
  Soldati]{soldati1}
{\sc \au{Soligo, G.}, \au{Roccon, A.} \& \au{Soldati, A.}}
  \yr{2019{\natexlab{{\em a\/}}}}  \at{Breakage, coalescence and size
  distribution of surfactant-laden droplets in turbulent flow}.  \jt{J. Fluid
  Mech.}  \bvol{881},  \pg{244--282}.

\bibitem[Soligo {\em et~al.\/}(2019{\natexlab{{\em b\/}}})Soligo, Roccon \&
  Soldati]{soldati3}
{\sc \au{Soligo, G.}, \au{Roccon, A.} \& \au{Soldati, A.}}
  \yr{2019{\natexlab{{\em b\/}}}}  \at{Mass-conservation-improved phase field
  methods for turbulent multiphase flow simulation}.  \jt{Acta Mech.}
  \bvol{230},  \pg{683--696}.

\bibitem[Stevens {\em et~al.\/}(2018)Stevens, Blass, Zhu, Verzicco \&
  Lohse]{richard}
{\sc \au{Stevens, R. J. A.~M.}, \au{Blass, A.}, \au{Zhu, X.}, \au{Verzicco, R.}
  \& \au{Lohse, D.}} \yr{2018}  \at{Turbulent thermal superstructures in
  {Rayleigh-B{\'{e}}nard} convection}.  \jt{Phys. Rev. Fluids}  \bvol{3},
  \pg{041501}.

\bibitem[Tackley(2000)]{science}
{\sc \au{Tackley, P.~J.}} \yr{2000}  \at{Mantle convection and plate tectonics:
  Toward an integrated physical and chemical theory}.  \jt{Science}
  \bvol{288},  \pg{2002--2007}.

\bibitem[{van der Poel} {\em et~al.\/}(2015){van der Poel},
  Ostilla-M{\'{o}}nico, Donners \& Verzicco]{cf15}
{\sc \au{{van der Poel}, E.~P.}, \au{Ostilla-M{\'{o}}nico, R.}, \au{Donners,
  J.} \& \au{Verzicco, R.}} \yr{2015}  \at{A pencil distributed finite
  difference code for strongly turbulent wall-bounded flows}.  \jt{Comput.
  Fluids}  \bvol{116},  \pg{10}.

\bibitem[{van der Poel} {\em et~al.\/}(2013){van der Poel}, Stevens \&
  Lohse]{2d}
{\sc \au{{van der Poel}, E.~P.}, \au{Stevens, R. J. A.~M.} \& \au{Lohse, D.}}
  \yr{2013}  \at{Comparison between two- and three-dimensional
  {Rayleigh-B{\'{e}}nard} convection}.  \jt{J. Fluid Mech.}  \bvol{736},
  \pg{177--194}.

\bibitem[Veron(2015)]{spray}
{\sc \au{Veron, F.}} \yr{2015}  \at{Ocean spray}.  \jt{Annu. Rev. Fluid Mech.}
  \bvol{47},  \pg{507–38}.

\bibitem[Verzicco \& Orlandi(1996)]{jcp96}
{\sc \au{Verzicco, R.} \& \au{Orlandi, P.}} \yr{1996}  \at{A finite-difference
  scheme for three-dimensional incompressible flows in cylindrical
  coordinates}.  \jt{J. Comput. Phys.}  \bvol{123},  \pg{402}.

\bibitem[Villermaux(2007)]{frag}
{\sc \au{Villermaux, E.}} \yr{2007}  \at{Fragmentation}.  \jt{Annu. Rev. Fluid
  Mech.}  \bvol{39},  \pg{419–46}.

\bibitem[Villermaux(2020)]{frag2}
{\sc \au{Villermaux, E.}} \yr{2020}  \at{Fragmentation versus cohesion}.
  \jt{J. Fluid Mech.}  \bvol{898},  \pg{P1}.

\bibitem[Villermaux \& Bossa(2009)]{rain}
{\sc \au{Villermaux, E.} \& \au{Bossa, B.}} \yr{2009}  \at{Single-drop
  fragmentation determines size distribution of raindrops}.  \jt{Nature Phys.}
  \bvol{5},  \pg{697--702}.

\bibitem[Wang {\em et~al.\/}(2020{\natexlab{{\em a\/}}})Wang, Chong, Stevens,
  Verzicco \& Lohse]{qi}
{\sc \au{Wang, Q.}, \au{Chong, K.~L.}, \au{Stevens, R. J. A.~M.}, \au{Verzicco,
  R.} \& \au{Lohse, D.}} \yr{2020{\natexlab{{\em a\/}}}}  \at{From zonal flow
  to convection rolls in {Rayleigh-B{\'{e}}nard} convection with free-slip
  plates}.  \jt{J. Fluid Mech.} 
  \bvol{905},
  \pg{A21}.

\bibitem[Wang {\em et~al.\/}(2020{\natexlab{{\em b\/}}})Wang, Verzicco, Lohse
  \& Shishkina]{qi2}
{\sc \au{Wang, Q.}, \au{Verzicco, R.}, \au{Lohse, D.} \& \au{Shishkina, O.}}
  \yr{2020{\natexlab{{\em b\/}}}}  \at{Multiple states in turbulent
  large-aspect ratio thermal convection: What determines the number of
  convection rolls?}  \jt{Phys. Rev. Lett.}  \bvol{125},  \pg{074501}.

\bibitem[Wang {\em et~al.\/}(2015)Wang, Shu, Shao, Wu \& Niu]{shu}
{\sc \au{Wang, Y.}, \au{Shu, C.}, \au{Shao, J.-Y.}, \au{Wu, J.} \& \au{Niu,
  X.-D.}} \yr{2015}  \at{A mass-conserved diffuse interface method and its
  application for incompressible multiphase flows with large density ratio}.
  \jt{J. Comput. Phys.}  \bvol{290},  \pg{336--351}.

\bibitem[Wang {\em et~al.\/}(2019)Wang, Mathai \& Sun]{chao}
{\sc \au{Wang, Z.}, \au{Mathai, V.} \& \au{Sun, C.}} \yr{2019}
  \at{Self-sustained biphasic catalytic turbulence}.  \jt{Nat. Commun.}
  \bvol{10},  \pg{3333}.

\bibitem[Xi {\em et~al.\/}(2004)Xi, Lam \& Xia]{xi04jfm}
{\sc \au{Xi, H.-D.}, \au{Lam, S.} \& \au{Xia, K.-Q.}} \yr{2004}  \at{From
  laminar plumes to organized flows: the onset of large-scale circulation in
  turbulent thermal convection}.  \jt{J. Fluid Mech.}  \bvol{503},
  \pg{47--56}.

\bibitem[Xie {\em et~al.\/}(2018)Xie, Ding \& Xia]{xie2}
{\sc \au{Xie, Y.-C.}, \au{Ding, G.~Y.} \& \au{Xia, K.-Q.}} \yr{2018}  \at{Flow
  topology transition via global bifurcation in thermally driven turbulence}.
  \jt{Phys. Rev. Lett.}  \bvol{120},  \pg{214501}.

\bibitem[Xie \& Xia(2013)]{xie}
{\sc \au{Xie, Y.-C.} \& \au{Xia, K.-Q.}} \yr{2013}  \at{Dynamics and flow
  coupling in two-layer turbulent thermal convection}.  \jt{J. Fluid Mech.}
  \bvol{728},  \pg{R1}.

\bibitem[Yoshida \& Hamano(2016)]{earth}
{\sc \au{Yoshida, M.} \& \au{Hamano, Y.}} \yr{2016}  \at{Numerical studies on
  the dynamics of two-layer {Rayleigh-B{\'{e}}nard} convection with an infinite
  {Prandtl} number and large viscosity contrasts}.  \jt{Phys. Fluids}
  \bvol{28},  \pg{116601}.

\bibitem[Yu {\em et~al.\/}(2020)Yu, Hendrickson \& Yue]{pdfjfm}
{\sc \au{Yu, X.}, \au{Hendrickson, K.} \& \au{Yue, D. K.~P.}} \yr{2020}
  \at{Scale separation and dependence of entrainment bubble-size distribution
  in free-surface turbulence}.  \jt{J. Fluid Mech.}  \bvol{885},  \pg{R2}.

\bibitem[Zhang {\em et~al.\/}(2017)Zhang, Zhou \& Sun]{zhou}
{\sc \au{Zhang, Y.}, \au{Zhou, Q.} \& \au{Sun, C.}} \yr{2017}  \at{Statistics
  of kinetic and thermal energy dissipation rates in two-dimensional turbulent
  {Rayleigh–B{\'{e}}nard} convection}.  \jt{J. Fluid Mech.}  \bvol{814},
  \pg{165--184}.

\bibitem[Zhu {\em et~al.\/}(2018{\natexlab{{\em a\/}}})Zhu, Mathai, Stevens,
  Verzicco \& Lohse]{zhu}
{\sc \au{Zhu, X.}, \au{Mathai, V.}, \au{Stevens, R. J. A.~M.}, \au{Verzicco,
  R.} \& \au{Lohse, D.}} \yr{2018{\natexlab{{\em a\/}}}}  \at{Transition to the
  ultimate regime in two-dimensional {Rayleigh-B{\'{e}}nard} convection}.
  \jt{Phys. Rev. Lett.}  \bvol{120},  \pg{144502}.

\bibitem[Zhu {\em et~al.\/}(2018{\natexlab{{\em b\/}}})Zhu, Verschoof, BaKolmogorov-Hinzeuis,
  Huisman, Verzicco, Sun \& Lohse]{zhu-tc}
{\sc \au{Zhu, X.}, \au{Verschoof, R.}, \au{BaKolmogorov-Hinzeuis, D.}, \au{Huisman, S.~G.},
  \au{Verzicco, R.}, \au{Sun, C.} \& \au{Lohse, D.}} \yr{2018{\natexlab{{\em
  b\/}}}}  \at{Wall roughness induces asymptotic ultimate turbulence}.
  \jt{Nature Phys.}  \bvol{14},  \pg{417--423}.

\bibitem[Zhu {\em et~al.\/}(2017)Zhu, Liu, Mu, Gao \& Ding]{zy}
{\sc \au{Zhu, Y.}, \au{Liu, H.-R.}, \au{Mu, K.}, \au{Gao, P.} \& \au{Ding, H.}}
  \yr{2017}  \at{Dynamics of drop impact onto a solid sphere: spreading and
  retraction}.  \jt{J. Fluid Mech.}  \bvol{824},  \pg{R3}.

\end{thebibliography}

\end{document}